\documentclass{article}

\usepackage[latin1]{inputenc}
\usepackage{times}
\usepackage{graphicx,amsmath,epsfig,epsf,psfrag}
\usepackage{mathptmx}
\usepackage{natbib}

\def \indi{1\!\mskip-2mu\mbox{\rmfamily I}}

\def \E{\hbox{I}\!\hbox{E}}

\begin{document}

%\begin{center}
{\noindent \bf \LARGE  A nonparametric model-based estimator for the cumulative
distribution function of a right censored variable in a finite population}
%\titlerunning{A model-based cdf estimator  of a right censored variable}

%\author{Sandrine Casanova,  Eve Leconte}
\begin{center}
{\sc  Sandrine Casanova,  Eve Leconte}
%\ead{sandrine.casanova@tse-fr.eu}
\bigskip

TSE (GREMAQ), 21, allée de Brienne,  31015 Toulouse Cedex 6, France\\

%\ead{eve.leconte@tse-fr.eu}

%\institute{Sandrine Casanova \at TSE (GREMAQ),  Universit\'e Toulouse 1 Capitole, \\
%  21, all\'ee de Brienne,  31015 TOULOUSE Cedex 6, France\\
%Tel.: +33-5-61128543\\
%Fax.: +33-5-61225563\\
E-mails: sandrine.casanova@tse-fr.eu, eve.leconte@tse-fr.eu
%\and
%Eve Leconte 
\end{center}
% \date{}

%\maketitle
\bigskip

{\bf Abstract:}
In survey analysis, the estimation of the cumulative distribution
function (cdf) is of great interest: it allows for instance to derive quantiles estimators or other non linear parameters derived from
the cdf.  We
consider the case where the response variable is a right censored duration variable. In this framework, the classical estimator 
of the cdf is the Kaplan-Meier estimator. As an alternative, we propose  a nonparametric model-based estimator of the
cdf in a finite population.  The new estimator
uses auxiliary information brought by a continuous covariate and is based on nonparametric median regression adapted to the censored
case. The bias and variance of the prediction error of the estimator are estimated by a bootstrap procedure adapted to censoring. 
The new estimator is compared  by model-based simulations to
the Kaplan-Meier estimator computed with the sampled individuals: a
significant gain in precision is brought by the new method whatever
the size of the sample and the censoring rate. Welfare  duration data are used to  illustrate
the new methodology.

{\bf Keywords:}
Cumulative distribution function, auxiliary
information, censored data, generalized Kaplan-Meier estimator,
nonparametric conditional median, bootstrap estimation.

\section{Introduction}

In survey sampling, the classical literature studies
estimation of totals or means but in many applications the parameters of interest are more complex: they can be quantiles (see
e.g. \citealp{Rue2004}) or other non linear parameters derived from
the cumulative distribution function (cdf) of the interest
variable. We consider the estimation of the cdf in a
finite population when the interest
variable is right censored.  This is the case when the interest
variable is a duration which is observed during a limited period of
time. For example, if we consider unemployment spells, individuals who
have not found a job at the end of the study have right censored
unemployment durations.  Notice that the censoring mechanism is
different from the nonresponse case: when the response variable of an
individual is censored, we know that  the duration for this individual
is  greater than the censoring time, whereas no information is
available for non respondents. Taking into account the partial
information brought by the censoring times improves the estimation.

To the best of our knowledge, there is no literature about the
estimation of the cdf in a finite population with right censored data.
This can be due to the fact that the censoring methodology has been
essentially developed in the medical field, where survey sampling is
not usual. Note that the classical cdf estimator  of a right censored variable in classical inference is the Kaplan-Meier estimator (\citealp{Kap1958}).

The estimation of the cdf in survey sampling has been widely studied
 in the absence of censoring (for a review, see by instance
Chapter 36 in \citealp{Pfe2009} and \citealp{Muk2001}). In a naive way,
the cdf is estimated by the empirical cdf computed on the sampled
individuals.  In the design-based approach, the conventional estimator of
the cdf is defined in a similar way but takes into account the
inclusion probabilities as for the Horvitz-Thompson estimator of a
total (see \citealp{Kuk1988}).  \citet{Rao1990} proposed a parametric
model-assisted estimator of the cdf and a nonparametric version of
this estimator was defined by \citet{Joh2008}.  In the following,
we will focus on model-based estimators.  In a parametric regression
framework, \citet{Cha1986} improve the estimation of the cdf by
predicting the response variable values of non sampled individuals
using auxiliary information brought by a covariate (this estimator will be denoted CD in the following).
\citet{Wan1996} construct a weighted average of the CD estimator and the estimator of \
 \citet{Rao1990} which performs better than the original estimators in terms of mean squared error.
Several variants of CD and
 \citet{Rao1990} estimators have been proposed (see Chapter 36 in \citealp{Pfe2009}).
 \citet{Dor1993}
define a nonparametric version of the CD estimator  and
study its asymptotic properties.

In section~\ref{sec2}, we propose a nonparametric model-based estimator
of the cdf for a finite population when the  variable of interest is right
censored.   The estimator
uses auxiliary information brought by a continuous covariate and is based on nonparametric median regression adapted to the censored
case.  In section~\ref{sec3}, the properties of the estimator  are discussed.
In section~\ref{sec4}, a bootstrap procedure to estimate the bias and variance of the prediction error is proposed. Section~\ref{sec5} compares the
performance of the new estimator to the naive Kaplan-Meier estimator computed with the sampled individuals by a model-based simulation study. An application to  a data set of welfare spells is presented in
section~\ref{sec6} and design-based simulations are performed in section~\ref{sec7}.  Some remarks are given in section~\ref{sec8}.

\section{Cdf estimation  of a censored variable in a finite population}
\label{sec2}

\subsection{Framework}

 In the following we will focus on model-based estimation so that the
 inclusion probabilities will not be used for the estimation.
 Therefore, we do not need to specify a sampling design.  However, to
 obtain consistent and efficient estimators, we need to assume that
 the sampling design is not informative (or ignorable), that is the
 same model holds for the sample and the population (see Introduction
 to Part 4 in \citealp{Pfe2009}).  Moreover we will propose
 a nonparametric estimator in order to reduce the risk of model
 misspecification.

Let us consider a finite population ${\cal P}$ with size $N$ and let $s$ be a sample
of ${\cal P}$ with size $n$.
The cdf of the interest
variable $T$ is therefore $F(t)=\displaystyle\frac{1}{N}\displaystyle\sum_{j \in
  {\cal P}}\indi(t_{j}\leq t)$ which can be partitioned into
\begin{equation}
F(t)=\displaystyle\frac{1}{N}\displaystyle\sum_{j \in
s}\indi(t_{j}\leq t)+\displaystyle\frac{1}{N}\displaystyle\sum_{j \in {\cal P}\setminus
s}\indi(t_{j}\leq t),
\label{decompositionpop}
\end{equation}
where $t_{j}$ is the value of the variable of interest measured for the
individual $j$ of the population ${\cal P}$.  Moreover, we suppose that
$t_{j}$ is a  non-negative value possibly right censored by a censoring time $c_{j}$.  So, on the sample $s$,
we observe $y_{j} = min(t_{j},c_{j})$ and
$\delta_{j}=\indi(t_{j}\leq c_{j})$.  We assume that auxiliary information
available on the whole population is given by a continuous covariate
$X$ and $x_j$ denotes the value of the covariate measured for the
individual $j$ of the population ${\cal P}$.

\subsection{A naive estimator of the cdf  $F$}

 It is well known that the empirical cdf does not provide a consistent estimator of the cdf in  the presence of censored data.
The cdf can be  consistently estimated by the Kaplan-Meier estimator (\citealp{Kap1958})
calculated on the sample $s$, which generalizes the empirical
cdf to the censored case.

Notice that the original Kaplan-Meier
estimator is undetermined after the last observed time
$y_{\left(n\right)}$ if this latter is censored. Therefore, to obtain
a distribution function, we will use the Efron's version (\citealp{Efr1967}) defined by:
\begin{equation}
\hat F_{\mbox{\scriptsize KM}}\left( t\right) =\left\{
\begin{array}{ll}
1 - \displaystyle \prod_{j\in s}\left\{ 1-\frac{1}{\displaystyle \sum_{r\in s}\displaystyle \indi \left(y_{r}\geq y_{j}\right)}
\right\} ^{
\displaystyle \indi
\left( y_{j}\leq t,\delta_{j}=1\right)} &
\mbox{if }\ t<y_{\left(n\right) }  \\[0.8cm]
1& \mbox{otherwise.}
\end{array}
\right.
\label{formuleKM}
\end{equation}
The Kaplan-Meier estimator is uniformly strongly consistent (see \citealp{Fol1980}) and under suitable regularity conditions, it converges weakly to a Gaussian process (see \citealp{Bre1974}).

\subsection{Cdf estimation using the prediction of the interest variable}

We propose  a model-based estimator of the cdf by estimating the two
terms of (\ref{decompositionpop}). Notice that the first term of
(\ref{decompositionpop}) is unknown because of right censoring and
must be estimated.  Since it can be written as:
\begin{equation}
\displaystyle\frac{1}{N}\displaystyle\sum_{j
\in s}\indi(t_{j}\leq t)= \displaystyle\frac{n}{N}\left(\frac{1}{n}\displaystyle\sum_{j
\in s}\indi(t_{j}\leq t)\right),
\label{premierterme}
\end{equation}
we recognize the cdf on the sample $s$ in the term in parenthesis.  This term can also be estimated by the Kaplan-Meier
estimator on the sample $s$.

In order to estimate nonparametrically the second term of
(\ref{decompositionpop}), we assume the superpopulation model:
\begin{equation}
\xi: \ \ t_{j}=m(x_{j})+\varepsilon_{j} \  \ (j \in {\cal P})
\label{supermodel}
\end{equation}
 where the
$\varepsilon_{j}$ are i.i.d. variables with cdf $G$ and $m(x_{j})$ is
the conditional median of $T$ given $X=x_{j}$.  
We have chosen to modelize the relationship between $t$ and $x$ by the conditional median instead of the classical conditional mean since the median  is  easier to estimate than  the mean in presence of right censored data.

As
$\E_{\xi}\left(\indi(t_{j} \leq t)\right)=P(t_{j} \leq
t)=G(t-m(x_{j}))$, a prediction of $\indi(t_{j} \leq t)$ can be
obtained by estimating $G(t-m(x_{j}))$.  Therefore, we first need to
estimate the conditional median $m(x_{j})$.  To this aim, we estimate
the conditional cdf of $T$ given $X=x$ with the generalized
Kaplan-Meier estimator (see \citealp{Ber1981}) on the sample $s$:
\begin{equation}
\hat F_{\mbox{\scriptsize GKM}}\left( t\mid x\right) =\left \{
\begin{array}{ll}
\displaystyle 1 - \prod_{j \in s}\left\{ 1-\frac{
B_{j}\left( x\right) }{\displaystyle \sum_{r\in s}B_{r}\left( x\right)\displaystyle \indi{\left(y_{r}\geq y_{j}\right)}
 }\right\} ^{
\displaystyle \indi
\left( y_{j}\leq t,\delta_{j}=1\right)}
& \mbox{if }\ t<y_{\left( n\right) }  \\[0.8cm]
1& \mbox{otherwise,}
\end{array}
\right.
\label{kapmeigen}
\end{equation}
where the $B_{j}(x)$ are Nadaraya-Watson type weights defined by:
$$
B_{j}\left( x\right) =\frac{\displaystyle K\left( \frac{x-X_{j}}{h_{X}}%
\right) }{\displaystyle \sum_{k \in s}K\left( \frac{x-X_{k}}{h_{X}}\right) }\cdot
\label{Bdab}
$$ $K$ is a kernel and $h_X$ denotes a suitable bandwidth.  It is easy
to check that $\hat F_{\mbox{\scriptsize GKM}}$ is a distribution function.
  Its uniform strong consistency  has been proved by \cite{Dab1989} and \cite{Gon1994}
  established a result of asymptotic normality with a norming factor
  of $\sqrt{nh_X}$.

As $\hat F_{\mbox{\scriptsize GKM}}$ is a step function with respect
to $t$, in order to estimate the conditional median by inversion, we
will use instead of $\hat F _{\mbox{\scriptsize GKM}}$ a
smoothed version in $t$ proposed by \citet{Lec2002}. Moreover, simulation studies have shown the gain brought by the
smoothing in $t$ in terms of the mean averaged squared error.
The proposed smoothed generalized Kaplan-Meier estimator is defined by:
\begin{equation}
{\hat F}_{\mbox{\scriptsize SGKM}}\left( t\mid x\right) =\sum_{l=1}^{{\#s^\dagger}+1}
\left( \hat F
_{\mbox{\scriptsize GKM}}\left( y_{l }^{\dagger }\mid x\right) -\hat F%
_{\mbox{\scriptsize GKM}}\left( y_{l-1 }^{\dagger }\mid x\right) \right) H\left(
\frac{t-y_{l }^{\dagger }}{h_T}\right),
\label{kapmeigensmooth}
\end{equation}
\noindent where $s^\dagger$ is the subset of the uncensored
individuals and the $\{y_l^{\dagger}, \ l=1,\ldots,\#s^\dagger\}$ denote the ordered times of
$s^\dagger$. In addition, we  use the following conventions: $y_0^{\dagger}=0$ and 
$y_{\#s^\dagger+1}^{\dagger}=y_{(n)}$. $H$ is an integrated kernel and $h_T$ is an appropriate
bandwidth.
Note that this smoothing  is  similar to the
classical kernel smoothing of the empirical cdf
by replacing the jumps $\frac{1}{n}$ of the em\-pi\-ri\-cal cdf by the jumps
of the generalized Kaplan-Meier estimator.
Thanks to the definitions
of $\widehat{F}_{\mbox{\tiny GKM}}$ and $H$, it is easy to check that
$\widehat{F}_{\mbox{\tiny SGKM}}\left( t\mid x\right)$ is a nondecreasing
function of $t$. The sum of the jumps is equal to $\widehat{F}
_{\mbox{\tiny GKM}}\left( y_{\left( n\right) }\mid x\right) $ which turns out to be
$1$ by formula (\ref{kapmeigen}). Therefore $\widehat{F}_{\mbox{\tiny SGKM}}(\cdot \mid x)$ is a
distribution function.
An estimator of the conditional median is then derived by numerical inversion as
$\hat m(x_{j})= \hat F_{\mbox{\scriptsize SGKM}}^{-1}(0.5\mid x_{j})$.

Now, let us return to the estimation of $G(t-m(x_{j}))$.  
As the residuals  
$\hat \varepsilon_{j}=y_j -\hat{m}(x_j), \ j \in s$  
may be right censored (obviously, $\hat \varepsilon_{j}$ is censored if $y_j$ is censored), a natural  estimator of the cdf $G$ of the errors  is the Kaplan-Meier estimator computed with the sampled residuals  $\hat \varepsilon_{j}$.
We denote this estimator $\hat G_{\mbox{\scriptsize{KM}}}$ and derive the following estimator of $F$:
\begin{equation}
\widehat F_{\mbox{\scriptsize M}}(t)=\displaystyle\frac{1}{N}\left(n\hat F_{\mbox{\scriptsize KM}}\left( t\right)+\displaystyle \sum_{j \in {\cal P}\setminus
s} \hat G_{KM} (t -\hat{m}(x_j))\right).
\label{formuleM}
\end{equation}

 It is
straightforward that $\hat F_{\mbox{\scriptsize{M}}}$ is a 
nondecreasing function.  Moreover, it tends to  1 when $t$ tends
to infinity. So, the proposed estimator is a genuine distribution function.
Note that this estimator, as well as the KM estimator,  has a natural extension in case of tied time values (see 2.4 of \citealp{Lec2002}).

\section{Properties of the new estimator}
\label{sec3}

\cite{Nas1995} have listed the properties required by a good estimator
of a cdf in a finite population.  The first one is
that the estimator should be a genuine cdf. This goal is achieved by
the estimator we have built.  Estimators of quantiles can then be
easily obtained by inverting the cdf estimator. 

 Another desirable property verified by the proposed estimator is the
 flexibility of the use of the auxiliary variable.  We assume in the
 above methodology that the auxiliary variable is continuous.
 However,
this  estimator
can be adapted to a discrete auxiliary variable by replacing the
generalized Kaplan-Meier estimator $\hat F _{\mbox{\scriptsize
    GKM}}\left( t\mid x_k\right)$ by the Kaplan-Meier estimator on the
subsample of individuals for whom the covariate is equal to $x_k$. In
addition to it, in the presence of several covariates, the auxiliary
information can be easily summarized by a univariate index computed for
instance performing a sliced inverse regression adapted to right censoring
\citep{Li1999}.

Moreover, the definition of the proposed estimator is relatively
automatic: as we use a nonparametric approach, no choices are required
in the specification of the model. The only choice is the
specification of the bandwidths which can be achieved by automatic
techniques such as cross-validation (see
section \ref{sec4}).

In a finite population, \cite{Dor1993} have shown that the nonparametric
version of the CD estimator is asymptotically model unbiased under some conditions concerning the bandwidth. They
also exhibit an asymptotic development for the variance of the
estimator leading to its consistency.  Because of the similarity of $\hat
F_{\mbox{\scriptsize M}}$ with the nonparametric version of the CD
estimator, we expect the new estimator to have similar asymptotic properties.
However these latter  can
not be obviously derived as an extension of the existing methodology because of the censorship.

Let us address the question of variance estimation.  An analytical
variance estimator for the CD estimator can be found in \cite{Wu2001}. They also develop a jacknife estimator
of the variance and show its design consistency.  \cite{Lom2004} have proposed to estimate by bootstrap the
bias, variance and prediction error of the nonparametric version of
the CD estimator
and they have shown the  consistency
of the used bootstrap estimator.
 Due to the presence of censoring and nonparametric
techniques which involve complex estimation procedures, an analytical
formula for the variance estimation of the new estimator has not yet been obtained.  However, in the next section,  we present an adaptation to the censored case of the bootstrap techniques of \cite{Lom2004}
in order to  estimate the bias and variance of the prediction errors of
the new estimator.

\section{Bootstrap estimation of the bias and variance of the prediction error}
\label{sec4}

Following \cite{Lom2004}, we use the argument proposed by \cite{Boo1994} which consists in estimating a characteristic of a finite population by averaging the values of the characteristics over booststrapped populations issued from the original sample.

Let us consider the original sample $(y_j,\delta_j,x_j)_{j \in s}$ with the superpopulation model $\xi$ (see (\ref{supermodel})), with the covariate $x$ known on the whole population  ${\cal P}$.
The adaptation of the \cite{Lom2004} method to the censored case leads to the bootstrap resampling
method  in three steps as follows:
\begin{enumerate}
\item Compute the residuals : $\hat \varepsilon_j=y_j - \hat m(x_j)$ as in section 2.3. and derive a smoothed Kaplan-Meier  estimator $\hat G_{\lambda}$ of $G$:
\begin{equation}
{\hat G}_{\lambda}(u) =\sum_{l=1}^{\#s^\dagger+1}
\left( \hat G
_{\mbox{\scriptsize KM}}\left(\hat \varepsilon_{\left( l\right) }^{\dagger }\right) -\hat G
_{\mbox{\scriptsize KM}}\left(\hat \varepsilon_{\left( l-1\right) }^{\dagger }\right) \right) H\left(
\frac{u-\hat \varepsilon_{\left( l\right) }^{\dagger }}{\lambda}\right)
\label{kmlisselambda}
\end{equation}
\noindent where $s^\dagger$ is the subset of the uncensored
individuals and the $\{\hat \varepsilon_{(l)}^{\dagger}, \ l=1,\ldots,\#s^\dagger\}\}$ denote the ordered residuals of
$s^\dagger$. In addition, we  use the following conventions: $\hat \varepsilon_0^{\dagger}=- \infty$ (${\hat G}_{\lambda}(\hat \varepsilon_0^{\dagger})=0$)  and 
$\hat \varepsilon_{\#s^\dagger+1}^{\dagger}=\hat \varepsilon_{(n)}$ (${\hat G}_{\lambda}(\varepsilon_{\#s^\dagger+1}^{\dagger})=1$). $H$ is an integrated kernel and $\lambda$ is an appropriate
bandwidth.

The bandwidths $h_T$ and $h_X$  have been chosen in a suitable grid of bandwidths so that they minimize a cross-validation criterion adapted to censoring defined   as follows:
\begin{equation}
CV=\sum_{j\in s^\dagger} |y_j-\hat m_{-j}(x_j)|
\label{CVcriterion}
\end{equation}
where $\hat m_{-j}(x_j)$ is the estimator of the conditional median based on $s$ minus the $j$th individual of $s^\dagger$. Note that we only use the uncensored durations in the CV criterion as the durations are not exactly known for censored observations.

As far as the choice of the $\lambda$ smoothing parameter is concerned, 
it has been chosen in a suitable grid  by cross-validation adapted to cdf estimation with censoring. Let $\lambda_0$ denote the value of $\lambda$ which minimizes the following criterion:
$$
\sum_{u \in {\cal G}_{\varepsilon}} \sum_{j\in s^\dagger} \left( \indi(\hat \varepsilon_j \leq u) - \hat G_{\lambda,-j}(u)\right)^2
$$
where  $\hat G_{\lambda,-j}(u)$ is the smoothed Kaplan-Meier estimator of $G$ based  on $s$ minus the $j$th individual of $s^\dagger$ and 
${\cal G}_{\varepsilon}$ is the grid of the 30 regularly spaced residuals in the range of the  $\hat \varepsilon_j$.
\item Generate  a $N$-membered bootstrap population ${\cal P^*}=(y_k^*,\delta_k^*,x_k)_{k \in {\cal P}}$  where 
$y_k^*=min(t_k^*,c_k^*)$ and
$\delta_k^*=\indi(t_k^*\leq c_k^*)$.
The bootstrapped event durations $t_k^*$ are obtained using the superpopulation model $\xi$ by $t_k^*=\hat m(x_k)+\varepsilon_{k}^*$, where the bootstrap errors $\varepsilon_{k}^*$ are generated according to $\hat G_{\lambda_0}$ by numerical inversion.
The bootstrapped censored durations $c_k^*$  have been obtained by inverting numerically the smoothed  Kaplan-Meier estimator of the cdf of the censored times from the original sample (known as the reverse Kaplan-Meier estimator).
\item Draw a sample $s^*$ of size $n$ from  ${\cal P^*}$ without replacement.
\end{enumerate}

Let $F^*(t)=\displaystyle\frac{1}{N}\displaystyle\sum_{k
\in {\cal P}}\indi(t_{k}^*\leq t)$ be the cdf of the $t^*$ variable.

The function $F^*$ can be estimated from the sample $s^*$, leading to an estimator denoted $\hat F^*$. 
Eq. (\ref{formuleKM}) (respectively  Eq. (\ref{formuleM})) gives the estimator $\widehat F^*_{\mbox{\scriptsize KM}}$ (respectively   $\widehat F^*_{\mbox{\scriptsize M}}$).
For computing time reasons, the bandwidths $h_T$ and $h_X$ have been chosen by data-driven techniques:  $h_{T}$ equals  30\% of the range of the $y$ and 
$h_X$ equal  30\% of the range of the $x$ in the bootsrapped sample.

Following \cite{Lom2004}, for an estimator $\hat F$ of $F$, we can estimate the bias $E(\widehat F(t)- F(t)| {\cal P})$ and the variance  $Var(\widehat F(t)- F(t)| {\cal P})$   of the prediction error 
using the predictors $E_*(E(\hat F^*(t)- F^*(t)| {\cal P^*}))$ and  $E_*(Var(\hat F^*(t)- F^*(t)| {\cal P^*}))$ respectively.
To approximate these predictors, according to step 2 and 3 of the previous procedure, we generate $B$ bootstrap populations denoted ${\cal P}^{*b}$$ (b=1,\dots,B)$ with size $N$  and from each one we draw $R$ samples with size $n$, denoted ${s}^{*br}$$ (r=1,\dots,R)$.
So we have the following approximations:
\[
\begin{array}{cc}
E_*(E(\hat F^*(t)- F^*(t)| {\cal P^*})) \approx \frac{1}{B}\frac{1}{R}\displaystyle\sum_{b=1}^B \displaystyle\sum_{r=1}^R [\widehat F ^{*br}(t)-F ^{*b}(t)]\\
E_*(Var(\hat F^*(t)- F^*(t)| {\cal P^*})) \approx \frac{1}{B}\frac{1}{R}\displaystyle\sum_{b=1}^B \displaystyle\sum_{r=1}^R [\widehat F ^{*br}(t)-\widehat F ^{*b}(t)]^2\\
\end{array}
\]
where $F ^{*b}$ is the cdf of the $b$th boostrap population, $\widehat F ^{*br}$ is the estimator or  $F ^{*b}$ computed from the $r$th sample of the $b$th bootstrapped population (with Eq. (\ref{formuleKM}) or Eq. (\ref{formuleM}))  and $\widehat F ^{*b}$ is the mean of the $R$ estimates  $\widehat F ^{*br}$ for a given $b$.

Moreover, following \cite{Lom2004}, a 100$(1-\alpha)$\% bootstrap confidence interval for $F$ 
can be obtained by 
\begin{equation}
CI\left[F(t)\right]^*=[\hat F(t)  - q^*_{1-\frac{\alpha}{2}}, \hat F(t) + q^*_{\frac{\alpha}{2}}]
\label{formuleIC}
\end{equation}
where  $\hat F(t)$ is computed from the original sample (with Eq. (\ref{formuleKM}) or Eq. (\ref{formuleM})) and  $q^*_{\alpha}$ is 
the 100$\alpha$-percentile of the bootstrap estimation of the function $H(u)=P(\hat F(t) - F(t) \leq u \mid {\cal P})$.

The original population ${\cal P}$ has been generated according to the
accelerated failure time model of subsection \ref{descrition_simu}
with HR=7.4, with $N=400$ and a censoring rate $\tau=25\%$.  $B=200$
bootstrapped populations have been generated and $R=1000$ samples have
been drawn from each population. The target cdf, its estimators as
well as  the bootstrap estimators have been computed on the grid ${\cal
  G}$ of the $K=30$ evaluation times $tt$ regularly spaced between the first
and the 99th percentiles of the $t$ values of the original sample.

\begin{figure}
\begin{center}
\includegraphics[angle=-90,width=\textwidth]{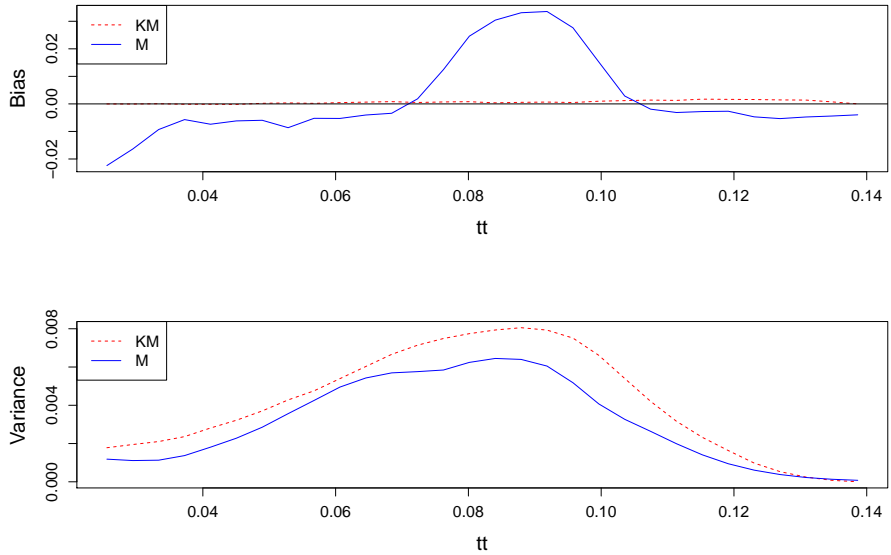}
\caption{Boostrap estimation of the biases and variances of the prediction error for the two estimators of the cdf ($B=200$, $R=1000$, $N=400$,  $\tau=25\%$). $tt$ denotes the time  values of the grid ${\cal G}$}
\label{bootstrapN400tau25biaisvar}
\end{center}
\end{figure}

Figure \ref{bootstrapN400tau25biaisvar} shows boostrap estimation of
the biases and variances of the prediction error for the two
cdf estimators $\hat F _{\mbox{\scriptsize KM}}$ and $\hat F
_{\mbox{\scriptsize M}}$. As expected, the bias of the prediction error is smaller for the estimator $\hat F_{\mbox{\scriptsize KM}}$ than for the estimator $\hat F_{\mbox{\scriptsize M}}$. In compensation,  the  
variance of the prediction error is weaker for the new estimator. 
The orders of magnitude of bias and variances are quite similar to those obtained by the model-based simulations (see section 5.2). 

\begin{figure}[ht]
\begin{center}
\includegraphics[angle=-90,width=\textwidth]{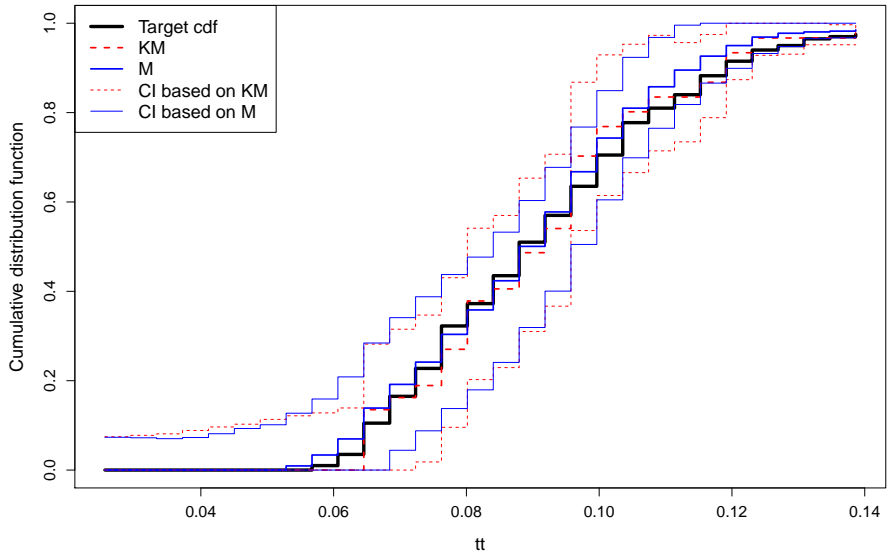}
\caption{Cdf estimators and the corresponding boostrap estimated  95\% confidence intervals  for $F$ ($B=200$, $R=1000$, $N=400$, $\tau=25\%$). $tt$ denotes the time  values of the grid ${\cal G}$.The target cdf has been computed from the original population ${\cal P}$}
\label{bootstrapN400tau25IC}
\end{center}
\end{figure}

Figure \ref{bootstrapN400tau25IC} presents the cdf $F$ 
with its two estimators $\hat F_{\mbox{\scriptsize KM}}$ and  $\hat F_{\mbox{\scriptsize M}}$ computed from the initial sample,
 as well as the 95\% bootstrap confidence intervals for $F$ based on formula (\ref{formuleIC}). 
The confidence interval based on $\hat F_{\mbox{\scriptsize M}}$ is more narrow than this based on $\hat F_{\mbox{\scriptsize KM}}$ for 83.3 \% of the t values of the grid.

\section{Model-based simulations}
\label{sec5}
\subsection{Description}
\label{descrition_simu}
We present a simulation study to compare the performances of the
two cdf estimators $\hat F_{\mbox{\scriptsize M}}$  and $\hat F_{\mbox{\scriptsize KM}}$, this latter being the naive
estimator of the cdf in presence of censoring. We have also derived estimators for the quartiles of the cdf.

At each iteration, a population of size $N$ ($N=200$ and $400$)
 has been generated according
to the accelerated failure time model $\log(t_{j})= -3 +0.2*
x_{j}+\sigma*u_{j}$ where the covariate $x_{j}$ is
uniformly distributed on $(1,4)$.  The error term $u_{j}$ follows an
extreme value distribution in order to obtain a Weibull distribution
for the $t_{j}$.  Note that this model is a proportional hazard model with a hazard ratio (HR) equal to $\exp(0.2/\sigma)$
which means that the ratio of the  hazard rates  of two individuals whose covariate $x$ differs from one unit is constant over time and  equal to $\exp(0.2/\sigma)$.
Two values of $\sigma$ (0.5 and 0.1) have been chosen leading to hazard ratios of 1.5 and 7.4, which correspond respectively to a weak and a strong relationship between the variable of interest and the auxiliary variable. $t_{j}$ is censored by $c_{j}$ where $c_{j}$ is
uniformly distributed on $(0,c)$, $c$ being chosen in order to obtain
$0 \%$, $10 \%$, $25 \%$ or $50 \%$ of censoring in the whole
population.   
At each iteration, we then draw a simple random sample without
replacement of size $n$=$N$/10. $S=1000$ iterations have been performed.

As far as the smoothing is concerned, we choose the triweight
kernel  $K\left( x\right) =\frac{35}{32}
\left( 1-x^{2}\right) ^{3}\indi_{\left( -1,1\right) }\left( x\right)$ rather than the more commonly used Epanechnikov kernel because the triweight kernel is twice differentiable at the boundaries of the interval  $\left( -1,1\right)$. So the resulting estimators will have the same degree of regularity.
For each iteration $s$, the bandwidths $h_T$ and $h_X$ have been chosen in a grid of bandwidths so that they minimize the averaged square error (ASE) criterion defined as:
$$
\mbox{ASE}(\hat{F}_{\mbox{\scriptsize M,s}})=\displaystyle\frac{1}{K}\sum_{i=1}^{K}\left(\hat{F}_{\mbox{\scriptsize M,s}}(tt_i)-F_s(tt_i)\right)^2.
$$
where the evaluation times $tt$ belong to the grid ${\cal G}$ of the $K=30$ regularly spaced
values of times between the 5th and the 95th percentiles of the distribution of $t$. Note that this grid is common to all the iterations.
The cdf $F_s$ is computed for iteration $s$
according to formula (\ref{decompositionpop}) using the true
$t_{j}$ times.

\subsection{Results}

The performances of the two estimators have been compared in terms of Monte Carlo bias, variance and mean squared error.  For  each
estimator $\hat F$, we compute the estimated bias
$$
\widehat{\mbox{B}}(\hat{F}(t))=\displaystyle\frac{1}{S}\sum_{s=1}^{S}\left(\hat{F}_s(t)-F_s(t)\right),
$$
the estimated variance
$$
\widehat{\mbox{Var}}(\hat{F}(t))=\displaystyle\frac{1}{S}\sum_{s=1}^{S}\left(\hat{F}_s(t)-\frac{1}{S}\sum_{s=1}^{S}\hat{F}_s(t)\right)^2
$$
and the estimated mean squared error (MSE):
$$
\widehat{\mbox{MSE}}(\hat{F}(t))=\displaystyle\frac{1}{S}\sum_{s=1}^{S}\left(\hat{F}_s(t)-F_s(t)\right)^2.$$

 Note that the usual relationship between the three above quantities
does not hold here since the $F_s$ function changes as the
population is generated at each iteration.  In practice, these
estimators have been computed on the grid ${\cal G}$ defined above.

The   MASE criteria (mean of the estimated MSE over ${\cal G}$) of the estimators have been computed and the ratios $MASE(\hat F_{\mbox{\scriptsize KM}})/MASE(\hat F_{\mbox{\scriptsize M}})$ are shown in
table~\ref{MASE} for two sample sizes, different censoring rates and two strengths of the relationship between the interest variable and the auxiliary variable. 

\begin{figure}[ht]
\begin{center}
\includegraphics[angle=-90,width=\textwidth]{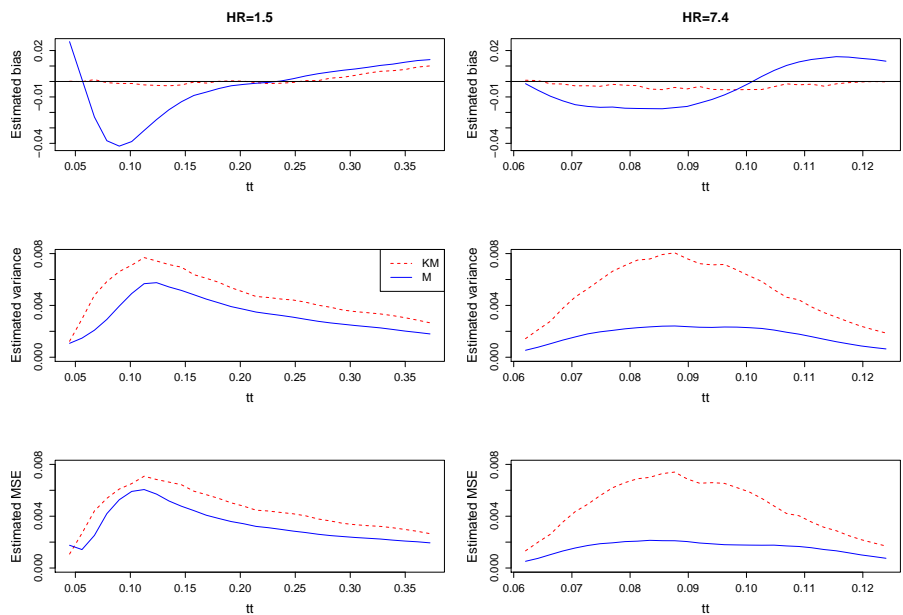}
\caption{Estimated biases, variances and MSE of the two estimators of the cdf for  N=400 individuals, a censoring rate $\tau=25\%$ and the two values of the hazard ratio HR. $tt$ denotes the  evalution times of the grid ${\cal G}$}
\label{biaisvarMSE}
\end{center}
\end{figure}

$\hat F_{\mbox{\scriptsize M}}$ performs  always better than $\hat F_{\mbox{\scriptsize KM}}$  with a maximal ratio of the MASE criteria equal to 3.03.
As expected, the gain brought by the auxiliary information is much higher when the relationship between the interest variable and the auxiliary variable is great: the ratios of the MASE are more than twice  greater when the hazard ratio
equals 7.4.
For both estimators, the simulations show that the  MASE criteria decrease with the sample size and increase with the censoring rate, but the ratios of the MASE criteria remain
almost the same for a given hazard ratio.
 
Figure~\ref{biaisvarMSE} shows the estimated bias, variance and MSE of
the two estimators of the cdf  for $N=400$ individuals
and a censoring rate $\tau=25\%$ for both hazard ratios. 
Notice that similar patterns are obtained for other sample sizes and
censoring rates. 
The new estimator has a greater bias than the estimator $\hat F_{\mbox{\scriptsize KM}}$ but it shows a smaller variance and  MSE for both values of the hazard ratio. As expected, when the relationship between the interest variable and the auxiliary is strong, the bias as well as variance and MSE are appreciably smaller.

\begin{table}
\caption{Estimated ratios $MASE(\hat F_{\mbox{\scriptsize KM}})/MASE(\hat F_{\mbox{\scriptsize M}})$. $n$ is the sample size, $\tau$ denotes the censoring ra\-te and HR is the hazard ratio of the proportional hazard model describing the relation between $t$ and $x$}
\label{MASE}
\begin{center}
\begin{tabular}{ccccccccc}
\hline\noalign{\smallskip}
&  \multicolumn{4}{c}{HR=1.5}&  \multicolumn{4}{c}{HR=7.4}\\
\noalign{\smallskip}\hline\noalign{\smallskip}
n  &  $\tau$=0\% & $\tau$=10\% & $\tau$=25\%  & $\tau$=50\%&  $\tau$=0\% & $\tau$=10\% & $\tau$=25\% & $\tau$=50\%\\
\noalign{\smallskip}\hline\noalign{\smallskip}
20 & 1.27&1.37 &  1.38    &   1.59&  3.03 &2.84 &2.85 & 2.81\\
40 & 1.27 & 1.33&1.34 & 1.46 & 2.88     & 2.96& 2.97& 2.85\\
\noalign{\smallskip}\hline
\end{tabular}
\end{center}
\end{table}

\begin{table}[ht]
\caption{Model-based simulation results for the estimation of the quartiles by the two estimators for a weak relationship ($HR=1.5$). Entries indicate relative biases, with square roots of the relavive mean squared errors  in parenthesis}
\label{simumed1}
\begin{center}
\begin{tabular}{cccccccccc}
\hline\noalign{\smallskip}
%& \multicolumn{8}{c}{Weak relationship ($HR=1.5$)}\\
&  \multicolumn{2}{c}{$\tau=0\%$}&  \multicolumn{2}{c}{$\tau=10\%$}& \multicolumn{2}{c}{$\tau=25\%$}&  \multicolumn{2}{c}{$\tau=50\%$}\\
&  \multicolumn{8}{c}{Target Q1: 0.080 }\\
\noalign{\smallskip}\hline\noalign{\smallskip}
n &  KM & M & KM & M & KM & M& KM & M \\
\noalign{\smallskip}\hline\noalign{\smallskip}
20& -0.004 &-0.021&0.023& -0.015&0.063 &-0.001&0.286& 0.049\\
 &(0.330) &(0.128) &(0.397)& (0.115) &(0.519)& (0.211) &(1.054)& (0.453)\\
40&-0.025& -0.024& -0.021& -0.019& -0.022& -0.017&-0.008 &-0.019\\
&(0.100) &(0.084)& (0.116)& (0.081)& (0.129)& (0.080)& (0.235)& (0.084)\\
\noalign{\smallskip}\hline\noalign{\smallskip}
&  \multicolumn{8}{c}{Target Q2: 0.108 }\\
\noalign{\smallskip}\hline\noalign{\smallskip}
n  &  KM & M & KM & M & KM & M& KM & M \\
\noalign{\smallskip}\hline\noalign{\smallskip}
20&  0.074 & 0.104&0.080  &0.107&0.107  &0.122&0.246& 0.133\\
 &(0.306)& (0.223)  &(0.316)& (0.228) &(0.389) &(0.263)&(0.748)& (0.364)\\
40&0.045 & 0.100 &0.063& 0.101 &0.074  &0.115 &0.061 & 0.099\\
&(0.187)& (0.195)& (0.199)& (0.197)& (0.209) &(0.212)& (0.233) &(0.205)\\
\noalign{\smallskip}\hline\noalign{\smallskip}
&  \multicolumn{8}{c}{Target Q3:  0.164}\\
\noalign{\smallskip}\hline\noalign{\smallskip}
n  &  KM & M & KM & M & KM & M& KM & M\\
\noalign{\smallskip}\hline\noalign{\smallskip}
20& 0.076 & 0.089&0.062 & 0.071&0.083  &0.088&0.071& 0.015\\
 & (0.295)& (0.228)  &(0.281)& (0.217)&(0.320) &(0.247)&(0.441)& (0.237)\\
40& 0.029 & 0.049&  0.047 & 0.046& 0.058 & 0.055&-0.036& -0.004\\
 &(0.191) &(0.162)& (0.198)& (0.165)& (0.223)& (0.186)& (0.165) &(0.136)\\
\noalign{\smallskip}\hline\noalign{\smallskip}
\end{tabular}
%\end{small}
\end{center}
\end{table}

\begin{table}[ht]
\caption{Model-based simulation results for the estimation of the quartiles by the two estimators for a strong relationship ($HR=7.4$). Entries indicate relative biases, with square roots of the relavive mean squared errors  in parenthesis}
\label{simumed2}
\begin{center}
\begin{tabular}{cccccccccc}
\hline\noalign{\smallskip}
&  \multicolumn{2}{c}{$\tau=0\%$}&  \multicolumn{2}{c}{$\tau=10\%$}& \multicolumn{2}{c}{$\tau=25\%$}&  \multicolumn{2}{c}{$\tau=50\%$}\\
&  \multicolumn{8}{c}{Target Q1: 0.075  }\\
\noalign{\smallskip}\hline\noalign{\smallskip}
n &  KM & M & KM & M & KM & M& KM & M\\
\noalign{\smallskip}\hline\noalign{\smallskip}
20&  -0.014 &0.011 &-0.015 &0.015&-0.014 &0.016&-0.030 & 0.012\\
&(0.092) &(0.037)&(0.095)& (0.038)&(0.107)& (0.044)&(0.139)& (0.057)\\
40& 0.001&  0.010 &0.003& 0.010&0.001 &0.012&-0.004& 0.013\\
&(0.046)& (0.026)&(0.053)& (0.027)&(0.059)& (0.030)&(0.077)& (0.034)\\
\noalign{\smallskip}\hline\noalign{\smallskip}
&  \multicolumn{8}{c}{Target Q2: 0.089 }\\
\noalign{\smallskip}\hline\noalign{\smallskip}
n  &  KM & M & KM & M & KM & M& KM & M\\
\noalign{\smallskip}\hline\noalign{\smallskip}
20& -0.003& 0.007&-0.004 &0.010&-0.005& 0.008&-0.010 & 0.006\\
&(0.073) &(0.043)&(0.071) &(0.043)&(0.076)& (0.044)&(0.089)& (0.051)\\
40&-0.006&  0.003&-0.004& 0.003&0.001 &0.006&-0.001& 0.004\\
& (0.056)& (0.039)&(0.056)& (0.040)&(0.060)& (0.041)&(0.069)& (0.045)\\
\noalign{\smallskip}\hline\noalign{\smallskip}
&  \multicolumn{8}{c}{Target Q3: 0.102}\\
\noalign{\smallskip}\hline\noalign{\smallskip}
n   &  KM & M & KM & M & KM & M& KM & M\\
\noalign{\smallskip}\hline\noalign{\smallskip}
20  & -0.016 &0.003& -0.015 &0.002&-0.020 &0.001&-0.035& -0.003\\
&(0.056) &(0.031)&(0.056)& (0.033)&(0.062) &(0.036)&(0.076)& (0.042)\\
40 &-0.007 &0.002&-0.002& 0.002&-0.003& 0.001&-0.012& 0.000\\
&(0.044) &(0.022)& (0.042) &(0.024)&(0.045) &(0.023)& (0.054)& (0.032)\\
\noalign{\smallskip}\hline
\end{tabular}
%\end{small}
\end{center}
\end{table}

The estimators of the quartiles have been obtained by numerical inversion of  the two cdf
estimators.
% according to the usual formula:
%$\widehat{\mbox{Q}}_{\alpha}= \mbox{inf}\{t : \hat{F}(t)\geq \alpha\}$
%for $\alpha$=0.25, 0.5 and 0.75. 
Tables \ref{simumed1} and  \ref{simumed2} show the
relatives biases and the square roots of the relavive mean squared
errors for the different sample sizes and censoring rates, for the two
hazard ratios.
The results are very similar to those obtained for the cdf estimation:
the  quartile estimator based on  $\hat F_{\mbox{\scriptsize KM}}$ has almost always a larger MSE than the quartile
estimator based on  $\hat F_{\mbox{\scriptsize M}}$.
As far as the relative bias is concerned, the estimator based on  $\hat F_{\mbox{\scriptsize M}}$ shows a better performance than the  estimator based on  $\hat F_{\mbox{\scriptsize KM}}$ in half of the cases. Notice that, when the auxiliary variable is strongly linked to the interest variable, the third quartile estimator based on  $\hat F_{\mbox{\scriptsize M}}$ always behaves better than the  estimator based on  $\hat F_{\mbox{\scriptsize KM}}$  in terms of bias and MSE criterion.
 Acccording to figure~\ref{biaisvarMSE}, this can be explained by
the fact that the curves of the biases of  $\hat F_{\mbox{\scriptsize M}}$ has  very small biases for the $t$ values close to the
third quartile.

\section{Example}
\label{sec6}

We analyse the  data
from the Survey of Income and Program Participation (SIPP) with the new method (see
\cite{Hu2012} for more details about the SIPP).
%Design-based simulations have been performed: they are based on data
%from the Survey of Income and Program Participation (SIPP) (see
%\cite{Hu2012} for more details).
We use the 1992 and 1993 SIPP
panels. Each individual is followed up during 36 months. We consider
the subsample of monoparental families who benefit from the Aid to
Families with Dependent Children program (AFDC). The $t$ variable of
interest is the length of time spent on welfare. For simplicity, only
the first welfare spell will be considered. The spell is
right-censored if it does not end before the family leaves the
panel. 520 spells have been recorded, among which 269 are right-censored,
leading to a censoring rate $\tau=$51.7\%. It has been found
in the literature that the benefit level is negatively and
significantly related to the probability of leaving welfare: in the
SIPP sample, a Cox model explaining the welfare duration by the
benefit level gives a hazard ratio of 0.999 (with a p-value of
0.0013).  Therefore we use the benefit level as auxiliary variable.

As we need to know the value of the auxiliary variable $x$ for the
whole population, we have to consider the above sample of 520 spells
as the fixed population ${\cal P}$, in which we draw a sample $s$
of size $n=40$ without replacement.
We compute the two cdf estimators $\hat F_{\mbox{\scriptsize KM}}$ and $\hat F_{\mbox{\scriptsize M}}$ based on the sample $s$ and the auxiliary variable $x$.
The bandwidths $h_T$ and $h_X$  have been chosen by cross-validation according to formula (\ref{CVcriterion}).
Bootstrap estimated 95\% confidence intervals for the cdf  based on the two estimators have been obtained by the procedure of section \ref{sec4} (see formula (\ref{formuleIC}).
As the variable of interest is censored in the considered population ${\cal P}$,
we cannot compute the true cdf. So, instead of the true cdf, we can use as a target cdf the Kaplan-Meier estimator $\hat F_N$ computed with
all the individuals of ${\cal P}$.
The estimators have been computed over the  grid  of the $K=30$ evaluation times $tt$ regularly spaced between the first
and the 99th percentiles of the $t$ values of the sample $s$.

\begin{figure}[ht]
\begin{center}
\includegraphics[angle=-90,width=\textwidth]{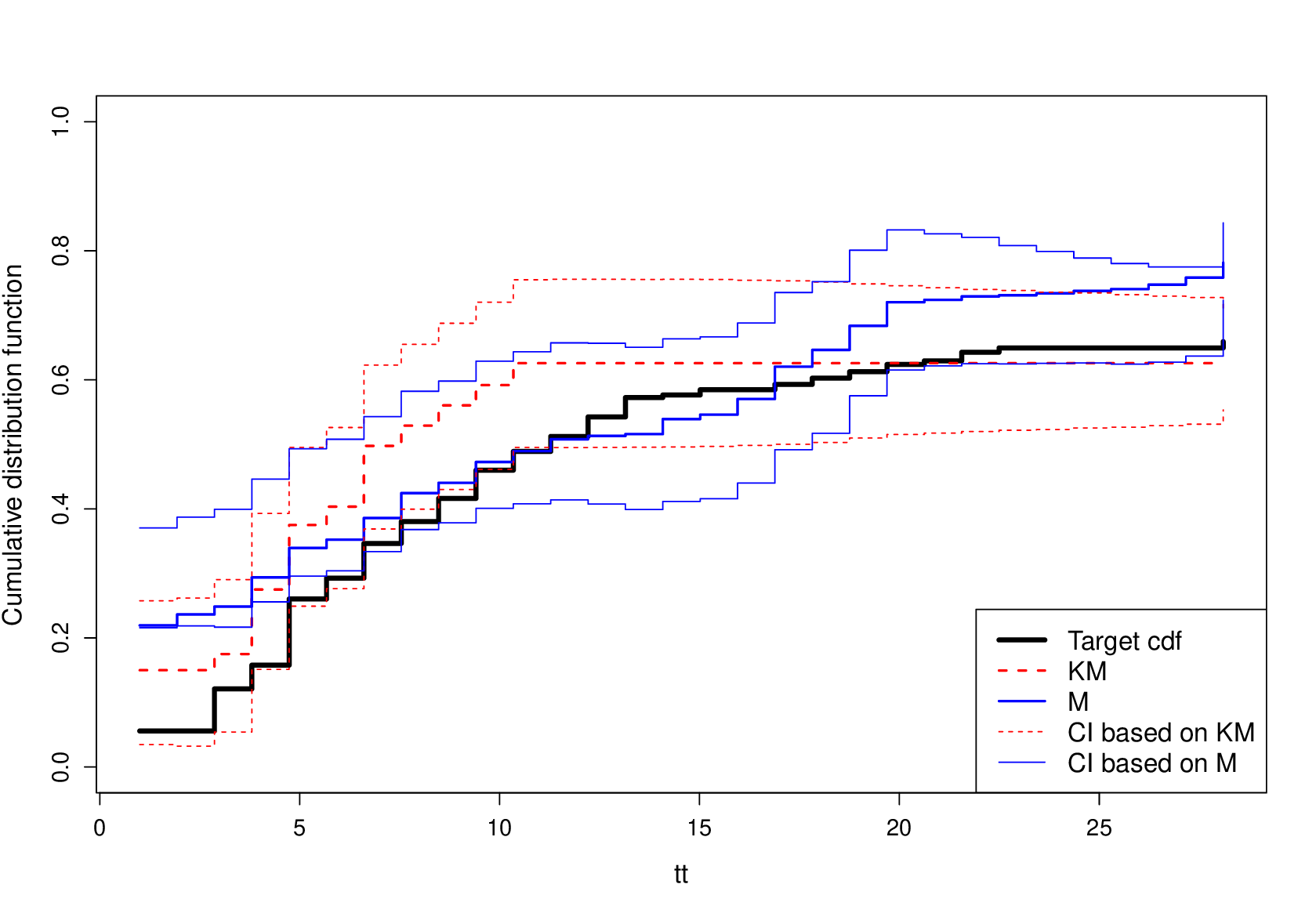}
\caption{Cdf estimators and the corresponding boostrap estimated  95\% confidence intervals  for $F$ ($B=200$, $R=1000$) based on a sample of size $n=40$. $tt$ denotes the time  values of the grid. The target cdf denotes the Kaplan-Meier estimator $\hat F_N$ computed on all the individuals of the SIPP sample}
\label{figure_exemple}
\end{center}
\end{figure}

Figure \ref{figure_exemple} presents the  two cdf estimators  $\hat F_{\mbox{\scriptsize KM}}$  and  $\hat F_{\mbox{\scriptsize M}}$ 
 as well as the corresponding 95\% bootstrap confidence intervals for $F$.
Note that the censoring rate of the drawn sample is 42.5\%.
We also plot as a reference the Kaplan-Meier estimator $\hat F_N$ computed on ${\cal P}$.
%Note that the two estimators are step functions and that the M estimator has more steps than the KM estimators
The confidence interval based on $\hat F_{\mbox{\scriptsize M}}$ is  more narrow than this based on $\hat F_{\mbox{\scriptsize KM}}$ for all the t values of the grid.
The  median welfare duration is estimated to  6.68 months by inverting $\hat F_{\mbox{\scriptsize KM}}$ and to 10.88 months by  inverting $\hat F_{\mbox{\scriptsize M}}$. This latter estimation is very close to the  estimation of the median welfare duration based on the Kaplan-Meier estimator $\hat F_N$, which equals 10.79 months.

\section{Design-based simulations}
\label{sec7}

Design-based simulations have been performed: they are based on the SIPP data 
 presented in the previous section.
To compare the two estimators, we consider  the SIPP sample of size 520 as a fixed population in which we randomly select $S=500$ samples of size 40 without replacement.
As in section \ref{sec6}, the true cdf $F$ can not be computed because of censoring.
Therefore  we use as a target cdf the Kaplan-Meier estimator $\hat F_N$ computed with
all the individuals of the SIPP sample.
For each iteration $s$, the bandwidths $h_T$ and $h_X$ have been chosen in a suitable grid of bandwidths so that they minimize the averaged square error (ASE) criterion defined as:
$$
\mbox{ASE}(\hat{F}_{\mbox{\scriptsize M,s}})=\displaystyle\frac{1}{K}\sum_{i=1}^{K}\left(\hat{F}_{\mbox{\scriptsize M,s}}(tt_i)-\hat F_N(tt_i)\right)^2,
$$
where  the evaluation times $tt$ belong to the grid ${\cal G'}$ of the $K=30$ regularly spaced values between the 5th and the 95th percentiles of the t values of the whole SIPP sample.

\begin{figure}
\begin{center}
\includegraphics[angle=-90,width=\textwidth]{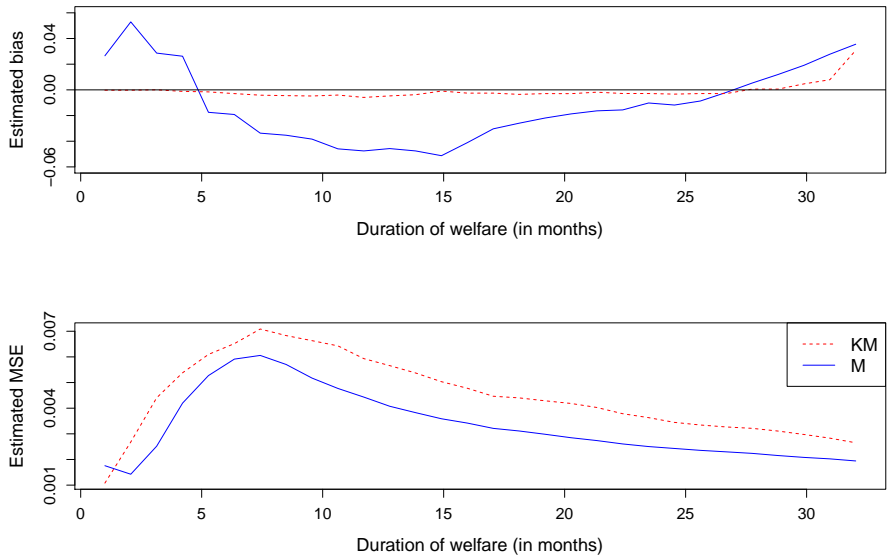}
\caption{Estimated bias and MSE of the cdf estimators for the welfare duration data}
\label{graphbiasmsewelfare}
\end{center}
\end{figure}

The ratio of the MASE criteria (mean of the ASE over the $S$ samples) of the estimator $\hat F_{\mbox{\scriptsize KM}}$ over the  estimator $\hat F_{\mbox{\scriptsize M}}$ is equal to 1.72, which shows clearly the gain brought by the new cdf estimator.
Table \ref{biasmsewelfare} presents
the relative bias and relative root mean squared errors of quartiles estimates of the distribution of the welfare spells.
The estimator $\hat F_{\mbox{\scriptsize M}}$  has the smallest relative bias except for the median and has always  the best performance  in terms of relative mean squared error.
Figure~\ref{graphbiasmsewelfare}  exhibits the estimated bias and mean squared errors (MSE)
of the two cdf estimators. 
As in the model-based simulations, the bias of $\hat F_{\mbox{\scriptsize KM}}$ is very close to zero.  On the other hand, $\hat F_{\mbox{\scriptsize M}}$ shows a more important bias but a substantially smaller mean squared error than  $\hat F_{\mbox{\scriptsize KM}}$.

\begin{table}
\caption{Relative biases and relative root mean squared errors  (in percentage) of estimates of quartiles of welfare durations ($n=40$)}
\label{biasmsewelfare}
\begin{center}
\begin{tabular}{lrrrl}
\hline\noalign{\smallskip}
        & \multicolumn{2}{l}{Relative bias} &   \multicolumn{2}{l}{Relative root MSE} \\
 Target quartile &         {\small KM}    &{\small M}   & {\small KM}       & {\small \quad \  \  M}  \\
\noalign{\smallskip}\hline\noalign{\smallskip}
$q_{0.25}$= 5.26 & 26.80 &3.18&  99.19& 36.09\\
$q_{0.50}$= 10.70 &  21.54 &27.82&51.33& 40.15\\
$q_{0.75}$=  22.60 & -14.15& 2.79 &  19.24& 12.22\\
\noalign{\smallskip}\hline
\end{tabular}
\end{center}
\end{table}

\section{Concluding remarks}
\label{sec8}

The simulations show the gain in precision by predicting the interest
variable for the non sampled individuals. Therefore it is worth using
the estimator  $\hat F_{\mbox{\scriptsize M}}$  instead of the Kaplan-Meier estimator  $\hat F_{\mbox{\scriptsize KM}}$ in a finite
population when auxiliary information is available.  

 According to formula
(\ref{formuleM}), it is obvious that $\hat F_{\mbox{\scriptsize M}}$ is a
step function with jumps among others things at the uncensored time values. As the
interest variable is continuous, we expect the cdf to be
continuous. So if desired, the cdf estimator 
$\hat F_{\mbox{\scriptsize M}}$ could be smoothed using for instance an integrated kernel as in formula (\ref{kapmeigensmooth}),
which would require another choice of bandwidth. 

The model-based approach is appropriate and will presumably lead to consistent
estimators when the sampling is not informative.  When a more complex
sampling method is used or when the sampling is informative, a
model-assisted approach which takes into account the sampling weights
would be more adapted. For instance, we can consider the model-assisted parametric cdf estimator of \citet{Rao1990} or its non-parametric version proposed by \citet{Dor1993} in the case of  simple random sampling. These estimators could be easily generalized to the censored case.

Note that, in panel surveys, nonresponse could be the
source of right censoring: in the design-based simulations of section
\ref{sec6}, an individual lost to follow-up who was still in welfare state
at his last interview is considered as censored. A methodology taking
into account the nonresponse could have been more adapted to this
case.

The proposed estimators are based on the generalized
Kaplan-Meier estimator of the conditional cdf.  Other estimators could
have been used. In particular, \citet{Kei2001} defined an estimator of
the conditional cdf which behaves better than the original Beran
estimator in the right tail of the distribution even under heavy censoring.
Alternatively, as proposed by \citet{Gan2005} in the censored case,
the conditional median could have been directly estimated by local linear
polynomials.

%\begin{acknowledgements}

%Thanks are due
%to the associate editor and the two referees for their helpful comments.

%\end{acknowledgements}

\bibliographystyle{spbasic}

%\bibliography{biblio}

\end{document}